\documentclass[12pt]{spieman}  % 12pt font required by SPIE;
\usepackage{amsmath,amsfonts,amssymb,aas_macros}
\usepackage{graphicx}
\usepackage{setspace}
\usepackage{tocloft}
\usepackage{siunitx}
\usepackage[symbol]{footmisc}

\newcommand{\mps}[1]{\qty{#1}{\meter\per\second}}

\newcommand{\daoa}{\ensuremath{\Delta\alpha/\alpha}}
\newcommand{\zdot}{\ensuremath{\dot{z}}}
\newcommand{\atom}[2]{#1{\sc \,#2}}
\newcommand{\cratiomath}{{\ensuremath{\rm \textsuperscript{12}C / \textsuperscript{13}C}}}

\title{Review of detector requirements: some challenges for the present}

\author[a,*]{Luca Pasquini}
\author[b,c,\dag]{Dinko Milakovi{\'c}}
\affil[a]{European Southern Observatory, Karl-Schwarzschild-Str.\, 2, Garching bei M\"unchen, Germany}
\affil[b]{Institute for Fundamental Physics of the Universe, Via Beirut 2, 34151 Trieste, Italy }
\affil[c]{INAF, Osservatorio Astronomico di Trieste, Via Tiepolo 11, 34131, Trieste, Italy}

\cftpagenumbersoff{figure}
\cftpagenumbersoff{table} 
\begin{document} 
\maketitle

\begin{abstract}
Astrophysics demands higher precision in measurements across photometry, spectroscopy, and astrometry. Several science cases necessitate not only precision but also a high level of accuracy. We highlight the challenges involved, particularly in achieving spectral fidelity, which refers to our ability to accurately replicate the input spectrum of an astrophysical source. Beyond wavelength calibration, this encompasses correcting observed spectra for atmospheric, telescope, and instrumental signatures. Elevating spectral fidelity opens avenues for addressing fundamental questions in physics and astrophysics. We  delve into specific science cases, critically analyzing the prerequisites for conducting crucial observations. Special attention is given to the requirements for spectrograph detectors, their calibrations and data reduction. Importantly, these considerations align closely with the needs of photometry and astrometry.

\end{abstract}

% Include a list of up to six keywords after the abstract

% Include email contact information for corresponding author
{\noindent \footnotesize\textbf{*}Luca Pasquini,  \linkable{lpasquin@eso.org} 
\noindent \footnotesize\textbf{\dag}Dinko Milakovi{\'c},  \linkable{dinko@milakovic.net} }

\begin{spacing}{2}   % use double spacing for rest of manuscript

\section{The quest for precision}
\label{sect:intro}  % \label{} allows reference to this section

Extensive datasets are essential for overcoming statistical noise and result interpretation is contingent upon systematic uncertainties remaining below the signal threshold. In other words, the effectiveness of large samples hinges upon the proper management of systematic uncertainties to yield meaningful results. While this work primarily delves into the requirements of optical high-resolution spectroscopy, similar requirements extend to other optical and near infra-red (NIR) instruments for astrometry or photometry. In fact, while one could na\"ively imagine that precision and big surveys lay on orthogonal axes of an observing plane, this is not the case. Some examples include space and ground based projects, such as 
\begin{enumerate}
\item{the \emph{Euclid} satellite\cite{Euclid2020A&A...635A.139E} requires that the variance of the residual shear systematics must be controlled to an accuracy of $\sigma^2_{sys}  \leq 10^{-7}$; }
\item{The centroid of each star observed by \emph{Gaia}\cite{Prusti2016A&A...595A...1G} is measured to a precision of about $10^{-3}$ per pixel by using many observations. This implies that the systematic effects must be controlled to the same level or better over the huge \emph{Gaia} focal plane  ;}
\item{MICADO\cite{Davies2021Msngr.182...17D} , the ESO-ELT imager will be assisted by adaptive optics (AO) and will require an astrometry precision better than $3\times10^{-2}$ per pixel; }
\item{Several spectrographs aim at detecting Earth mass planets in the habitable zone of solar stars via radial velocities (RV) measurements. The orbit of Earth induces a RV wobble of less than 10 cms$^{-1}$/yr in the solar spectrum, while the typical pixel size of high-resolution spectrographs is in the 500-1000 ms$^{-1}$ range, i.e.\ $>$5000 times larger than the signal. }

\end{enumerate}

\subsection{Spectral Fidelity}
Spectral fidelity is our capability to accurately reproduce the input spectrum of an astrophysical source from sensor output. In addition to the wavelength calibration, it includes our ability to correct the observed spectrum for the atmosphere, telescope and instrumental signatures. By bringing spectral fidelity to a new height, many questions can be addressed, as discussed at the ‘Spectral Fidelity’ conference\footnote[3]{https://www.eso.org/sci/meetings/2023/fidelity.html} in September 2023 and as discussed here. 

One intriguing aspect of astronomical spectroscopy is that the results from two instruments (and sometimes even from two authors analysing the same data) are not reproducible, and not compatible. Rather fundamental examples include the debate on the photospheric chemical composition of the Sun\cite{Bergemann2021MNRAS.508.2236B} or the ongoing investigation on the variability of the fine structure constant\cite{Webb2024MNRAS.528.6550W,Lee2021MNRAS.507...27L_nonunique} . Some difficulties are intrinsic to the complex nature of the astrophysical objects, and to the limitation that we cannot reproduce these objects in a laboratory, but others arise from the limited accuracy of our spectra: the spectra of the same object acquired with different spectrographs are not consistent at the necessary level. In this study, we intentionally refrained from discussing the astrophysical challenges, instead focusing on the observational aspects. This rationale is rooted in the belief that establishing a solid foundation in observations and data reduction is essential before delving into astrophysical complexities.

We discuss several science cases that would benefit from improvements to spectral fidelity. For example, some of the most demanding projects related to spectroscopy of the intergalactic medium (IGM) are (1) directly measuring the acceleration of the universe, (2) constraining any variation in the values of fundamental physical constants, (3) studying isotopic abundances in the early universe (such as Deuterium and Carbon). Similarly, the most demanding stellar spectroscopy projects relate to (1) discovering new planetary systems and characterizing exoplanet atmospheres, (2) making accurate chemical element abundance measurements, including isotopic ratios of various elements. The {\it legacy} aspect of spectral fidelity is important as well. We can image of transformational projects combining the spectra obtained with different instruments, or making extensive use of archival data in decades from now.  

In the past 20 years enormous steps forward were made in instrumentation, with the adoption of spectrographs' fibre injection optimized for scrambling capabilities, simultaneous wavelength calibration and use of temperature and pressure stabilized instruments. The last generation of spectrographs, such as ESPRESSO\cite{Pepe2021A&A...645A..96P} at the Very Large Telescope (VLT), basically fulfills all the requirements for high precision and accurate work. Two important, and not yet fully achieved steps, are the use a suitable metrology (calibration) system, one that can allow the full exploitation of the instrument, and a data reduction that subtracts all instrumental signatures. Achieving these steps will certainly open up a new discovery space and produce important scientific discoveries.

\section{Applications requiring high spectral fidelity}

\begin{figure}
\begin{center}
\begin{tabular}{c}
\includegraphics[height=7.5cm]{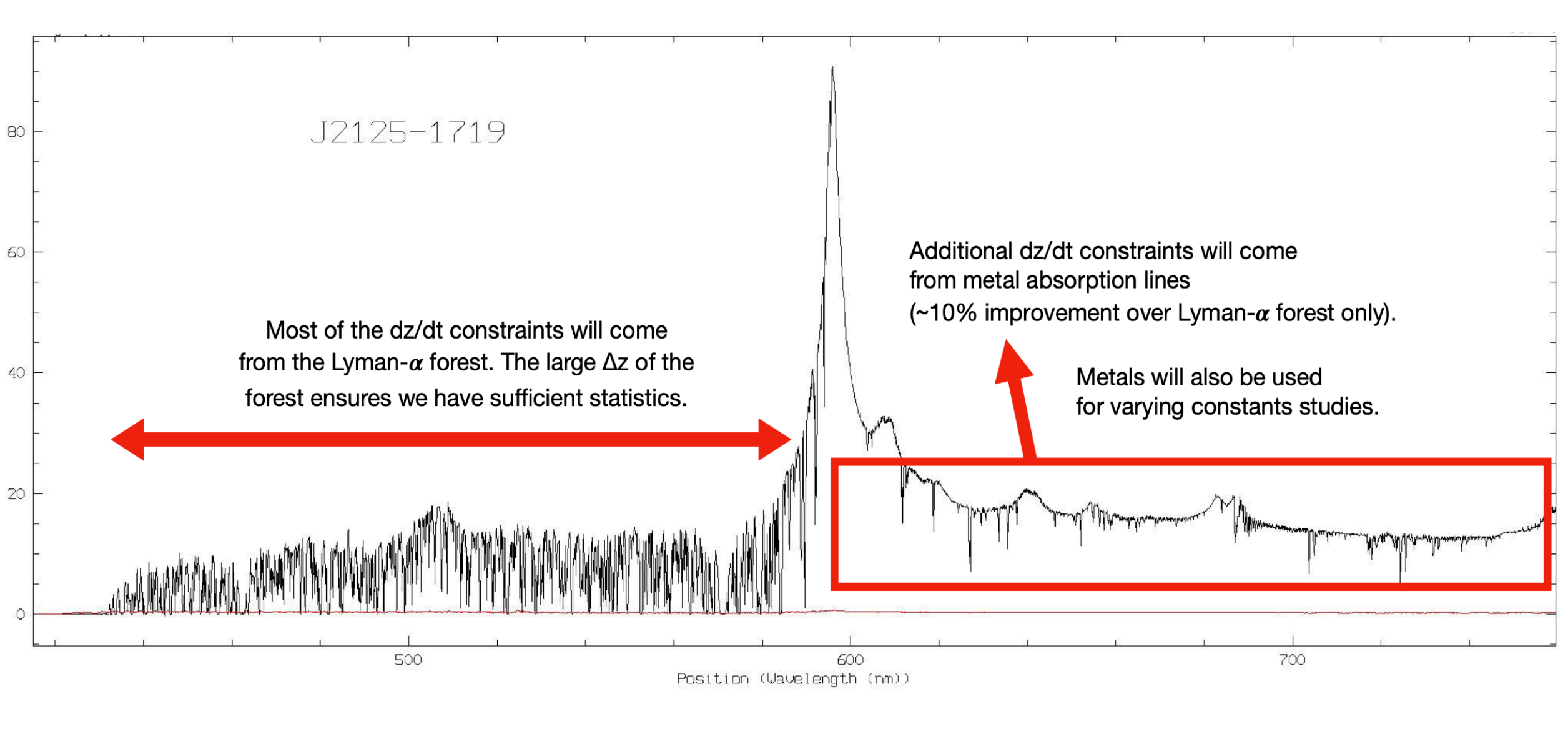}
\end{tabular}
\end{center}
\caption [Spectrum of QSO J2125$-$1719]
{ \label{fig:qso}
Spectrum of J2125$-$1719, a bright $z\sim4$ QSO observed recently by the QUBRICS survey, showing the portions of the spectrum used for the different fundamental physics and cosmology tests. } 
\end{figure}

Fig~\ref{fig:qso} shows the intermediate resolution spectrum of a bright z$\sim$4 quasi-stellar object (QSO or quasar) recently discovered by Schindler et al.\cite{Schindler2021ApJ...906...12S} and observed by the QUBRICS survey\cite{Cristiani2023MNRAS.522.2019C}. The dense hydrogen Ly$\alpha$ forest at wavelengths shorter than the Ly$\alpha$ emission is ideal for performing the redshift drift measurement\cite{Sandage1962ApJ...136..319S,McVittie1962ApJ...136..334M,Loeb1998ApJ...499L.111L,Liske2008MNRAS.386.1192L}, while the heavy element absorption lines (primarily present at wavelengths longer than the Ly$\alpha$ emission), are typically used for testing the variability of the fine structure constant $\alpha$. 

\subsection{The redshift drift measurement}\label{sec:redshift_drift}
The physical explanation as to why universal expansion appears to be accelerating in the past few billion years is unclear. The cosmological constant $\Lambda$ originally introduced by Einstein to explain a static non-expanding universe can explain this late acceleration stage very well, but its value corresponding to an equivalent energy density similar to that of dark matter appears rather strange, considering that the ratio between the two could have taken any value. Modification of General Relativity and the presence of ``dark energy'' are thus under vivid study by theoretical and observational astrophysicists. 

Probably the best way to probe the nature of the acceleration is to accurately determine the expansion history of the universe. Observables that depend on the expansion history include distances and the linear growth of density perturbations, making surveys of type Ia supernovae, weak gravitational lensing, and the signature of the sound horizon in the galaxy power spectrum excellent probes of the acceleration. In practice, however, extracting information on the expansion history from these surveys requires a prior on the spatial curvature, a detailed theoretical understanding of the linear growth of density perturbations and hence a specific cosmological model. A model-independent approach to determining the expansion history is thus highly desirable.

Sandage\cite{Sandage1962ApJ...136..319S} and McVittie\cite{McVittie1962ApJ...136..334M} first discussed the slow temporal drift of the redshifts of cosmologically distant objects due to the change of the expansion of the universe. If observed, the redshift drift rate $\zdot = {\rm d}z / {\rm d} t$, would constitute direct evidence of the universal deceleration or acceleration between redshift $z$ and today, offering a model-independent measurement of the expansion history and uniquely probing the global dynamics of the space-time metric. The Ly$\alpha$ lines seen towards bright QSOs are numerous, ubiquitous, reasonably narrow (allowing for precise centroiding), and their peculiar accelerations are expected to be random (such that they can be averaged over). They can therefore be used to measure $\zdot$ over a large range of cosmic history. 

Liske et al.\cite{Liske2008MNRAS.386.1192L} addressed the ``in-principle'' feasibility of measuring the redshift drift of the Ly$\alpha$ forest over $\sim20$ years using a 40-m class optical telescope. Because the drift rate is expected to be extremely small, $\sim$ 6 cms$^{-1}$ per decade at $z=4$ and $\sim$ 3 cms$^{-1}$ per decade at $z=3$, a successful measurement presumes several conditions to be met. A 3$\sigma$ detection from observations spanning 20 years would require (1) an average uncertainty (over some redshift range) on Ly$\alpha$ clouds' redshifts to be 2-3 cms$^{-1}$ and (2) measuring instrumental drifts with the same precision over the entire time period, such that their impact on the scientific analysis can be removed. To achieve the first condition, the signal-to-noise ratio (S/N) of all collected observations should be higher than 10000 per pixel \cite{Liske2008MNRAS.386.1192L}. However there are only a few currently known QSOs that are bright enough to reach such high S/N, even with an extremely large telescope (ELT). Achieving the second requirement requires a significantly better understanding of the instrument's detector and better calibration procedures, as discussed in Section \ref{sec:reaching_spectral_fidelity}. Uncertainties in wavelength calibration currently dominate the instrumental uncertainty budget. On HARPS\cite{Mayor2003Msngr.114...20M}, practical limits of using astronomical laser frequency combs (``astrocombs'') for absolute wavelength calibration were explored with an experiment that saw two independent combs operated simultaneously for several days. Surprisingly, the two astrocombs disagreed in what is the wavelength calibration zero-point of the instrument by 45 cms$^{-1}$, even after considering flux-dependent effects on the detector \cite{Milakovic2020MNRAS.493.3997M}. The nominal precision of each astrocomb 3 cms$^{-1}$, making this a $15\sigma$ discrepancy. Should instruments used for this experiment have a systematic of this magnitude in the determination of their zero-point, detecting the redshift drift with $3\sigma$ significance would take $>$200 years.

Although it is not trivial to translate all of the above (S/N per pixel, uncertainties on line centroid measurements, tracking drifts in the wavelength calibration) into detector requirements, it is clear that these are very demanding performances, also considering that such a high S/N will be obtained by adding up many observations at a much lower S/N. It will certainly require a detailed mapping of the pixel response function (with sub-percent accuracy across a wide dynamic range in flux), measuring the charge transfer efficiency of all detector pixels (unless each pixel is read out individually), identifying and, if possible, calibrating any problematic pixels, and properly wavelength calibrating the spectra. It will also be necessary to identify issues related to persistence, read-out electronics, thermal noise, brighter-fatter effect, etc. Not related to detectors, but equally important, will be to characterise the point-spread function (PSF) shape as a function of position on the detector, flux level, and time.

Any complications are likely to extend the duration of the experiment. It is, therefore, of paramount importance to be able to calibrate the spectrograph and its detector as accurately as possible from the very start of observations to minimise the risk of significant delays or even failure. It is very likely, and perhaps even advisable, that multiple spectrographs should be used to observe the same objects over several decades, for cross-check and continuity. For the cross-instrument comparison to work, the spectrographs' wavelength calibration must be known with an absolute accuracy that matches the final uncertainty on Ly$\alpha$ line centroid measurements in the data due to photon noise. Should this be achieved, the spectrograph would not need to be stable at  cms$^{-1}$ level for longer than the duration of a typical observation ($\sim 1$h).

\subsection{Fine structure constant variability}

The mathematical formalism behind the laws of physics (as we currently understand them) requires a set of $\approx 22$ free parameters, or fundamental constants. Together, these constants determine the fundamental properties of our universe (see, e.g., the reviews by Uzan\cite{Uzan2011LRR....14....2U,Uzan2024arXiv241007281U}) and we have (implicitly) assumed that their values are indeed constant and independent from each other. Variations of the fine structure constant, $\alpha=e^2/(\hbar c 4\pi\varepsilon_0)$ (SI), are on the other hand predicted by many theories aiming to provide an explanation for dark matter and dark energy \cite{Bekenstein1982PhRvD..25.1527B,Barrow2012PhRvD..85b3514B,Shaw2005PhRvD..71f3525S,Sandvik2002PhRvL..88c1302B,Mota2004PhLB..581..141M} . Any potential variation in $\alpha$ manifests as a perturbation of energy levels of atoms and results in small frequency shifts of atomic transitions with respect to their measured laboratory values, with different transitions having different sensitivities to $\alpha$ variation. For the same relative change in $\alpha$, different transitions thus exhibit shifts of different amplitudes and signs. Very stringent constraints for non-variability come from atomic clock measurements on Earth, albeit over short time periods ($\sim1$ year) \cite{Rosenband2008Sci...319.1808R} . 

Metal transitions in QSO spectra provide a way to measure the value of $\alpha$ in the early universe. The Many Multiplet (MM) method \cite{Dzuba1999PhRvA..59..230D,Dzuba1999PhRvL..82..888D,Webb1999PhRvL..82..884W} , leverages the fact that ground state transitions are generally the most sensitive, and that a typical quasar spectrum contains ground transitions of several different atomic species. For example, a variation of one part per million (ppm) in $\daoa$ introduces a $\approx\mps{15}$ shift between line centers of \atom{Fe}{ii} $\lambda 2344$ and \atom{Mg}{ii} $\lambda2796$. Because the method relies on comparing relative shifts between a set of transitions with (rest-frame) separations up to $\sim\qty{1500}{\angstrom}$, wavelength calibration distortions must be smaller than $\sim\mps{5}$ across the entire relevant wavelength range. 

The application of the MM method to 298 QSO absorption systems observed with UVES \cite{Dekker2000SPIE.4008..534D} at the VLT and with HIRES \cite{Vogt1994SPIE.2198..362V} at the Keck telescope provided first evidence for temporal and spatial variations of $\alpha$, at $\daoa\sim10^{-5}$ level \cite{Webb2001PhRvL..87i1301W,Webb2003Ap&SS.283..565W,Murphy2003MNRAS.345..609M_measurements,Webb2011PhRvL.107s1101W,King2012MNRAS.422.3370K} .  
However, a cross-check of the Th-Ar based wavelength calibration using asteroid observations revealed the presence of long-range wavelength distortions with slopes as large as \mps{200} per \qty{1000}{\angstrom} in both UVES and HIRES \cite{Griest2010ApJ...708..158G,Whitmore2010ApJ...723...89W,Whitmore2015MNRAS.447..446W,Rahmani2013MNRAS.435..861R} . 
%Na\"ively, this corresponds to an error of 30 ppm using the above rule of thumb but the actual impact on the measurement is much smaller \cite{Dumont2017MNRAS.468.1568D} . 
Whilst these distortions can be minimised with extra calibration frames \cite{Molaro2008A&A...481..559M,Rahmani2013MNRAS.435..861R} and their impact can propagated into the total uncertainty within the spectral modelling process \cite{Dumont2017MNRAS.468.1568D} , wavelength scale distortions remained difficult to control. It was only with application of astrocombs to HARPS\cite{Mayor2003Msngr.114...20M} and ESPRESSO that systematic uncertainties associated with wavelength calibration distortions became smaller than other sources of uncertainty \cite{Milakovic2020MNRAS.493.3997M,Milakovic2021MNRAS.500....1M,Milakovic2024A&A...684A..38M,Schmidt2021A&A...646A.144S,Milakovic2023MmSAI..94b.270M} . 

\begin{figure}
\begin{center}
%\begin{tabular}{c}
\includegraphics[height=7.5cm]{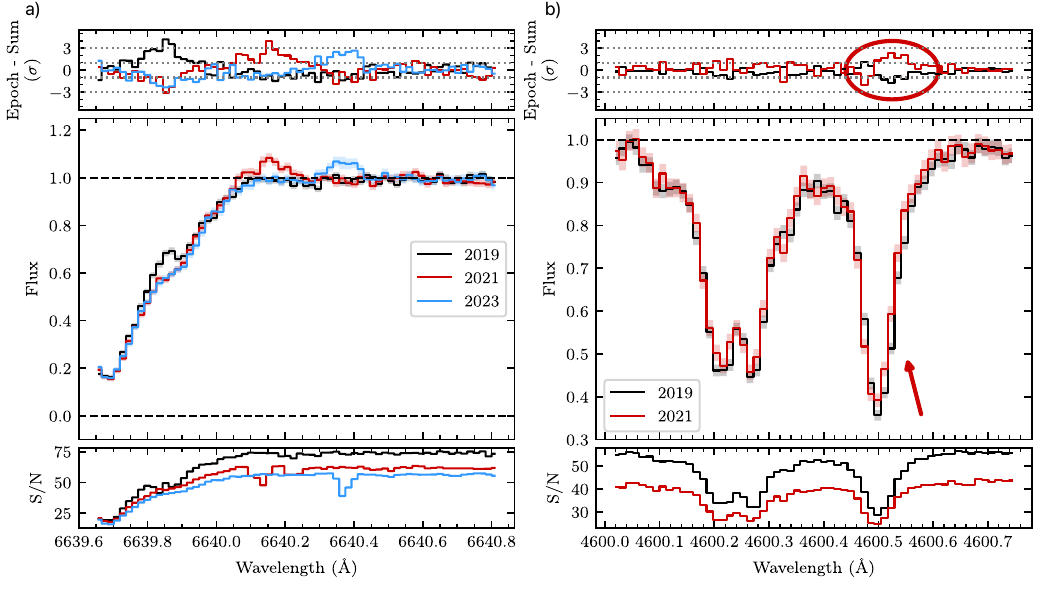}
%\end{tabular}
\end{center}
\caption [Spectra of QSO J1333$+$1649 showing the effect of warm detector pixels]
{ \label{fig:espresso_warm_pixels}
Example of the effect unmasked warm detector pixels have on extracted spectra. Both panels show small spectral segments of ESPRESSO spectra of QSO J1333$+$1649 observed in three epochs: 2019 (black), 2021 (red), and 2023 (blue). Left panels (labelled a) show a part of \atom{Fe}{ii} $\lambda 2382$ at $z_{abs}=1.786$ and the right panels (labelled b) show \atom{C}{i} $\lambda1656$ at $z_{abs}=1.776$. Errors are shown as shaded regions around the flux (full lines). Clusters of warm detector pixel are seen as excess flux spanning several consecutive pixels. These were not correctly identified by the data reduction pipeline so have propagated into the final scientific spectrum. The top panel shows the difference between each epoch and the sum of the plotted spectra, normalised by the total error. Individual pixels within the cluster are discrepant at  $1\sigma$ to $3\sigma$ range (thin dotted black lines). Collectively, the clusters are $>5\sigma$ discrepant and prevent a robust scientific analysis. The S/N per pixel is shown in the bottom panel, and is seen to drop at the locations of the clusters because of outlier search performed during the co-addition process. However, only a small fraction of warm pixels could be removed by this procedure. The colour scheme is the same in all panels. } 
\end{figure}

In fact, advancements implemented in ESPRESSO removed many, previously dominant, instrumental sources of uncertainty and shifted focus on effects that were previously less important. For example, clusters of ``warm'' pixels were found in ESPRESSO observations of faint objects. The left panel of Fig \ref{fig:espresso_warm_pixels} shows a small spectral segment of the QSO J1333$+$1649 observed by ESPRESSO in three different epochs: 2019, 2021, and 2023. In the Figure, a cluster spanning 5-10 pixels appears at different locations in the wavelength calibrated spectrum due to different barycentric Earth radial velocity corrections for the three epochs. So far, these artefacts have only been observed in long ($\sim 1$h) exposures typical for quasar observations and in dark frames because it takes time for the electrons to accumulate in the potential wells of affected pixels. Consequently, conventional detrending techniques based on flat-field frames are unable to remove them. In a single quasar exposure, flux within the warm pixel cluster can be up to 10\% larger than the flux in surrounding pixels, albeit most warm pixels are consistent with noise ($<3\sigma$) due to overall low level of recorded flux. However, if not removed, spectral co-addition procedures increase the statistical significance of the artefacts to $>5\sigma$, as seen in this example. Identification of the warm pixels in raw frames is tricky and they are often missed by standard pipeline procedures. Modifying the pipeline's clipping threshold to identify the warm pixel clusters also removes good data. 

\subsection{Isotopic ratios in the IGM}
\begin{figure}
\begin{center}
\begin{tabular}{c}
\includegraphics[height=7.5cm]{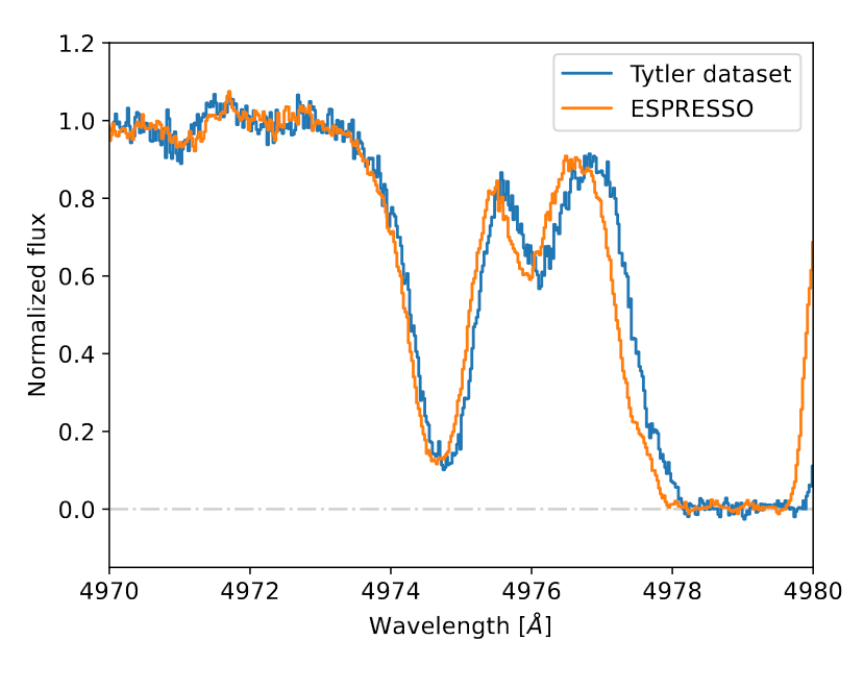}
\end{tabular}
\end{center}
\caption [Spectra of QSO PKS1937$-$101 in the D line region]
{ \label{fig:deu}
Spectra of QSO PKS1937$-$101 in the D line region taken with ESPRESSO and HIRES. The difference between the spectra cannot be explained by a simple velocity shift, but shows small scale wavelength calibration distortions. Consequently, a joint analysis of the HIRES, ESPRESSO, and UVES spectra did not improve the D/H measurement. Taken from Guarneri et al.\cite{Guarneri2024MNRAS.529..839G}.    } 
\end{figure}

The quantity of Deuterium formed during Big Bang nucleosynthesis (BBN) can be determined from observations of the Ly$\alpha$ transitions of metal poor, and hence pristine, IGM. The deuterium Lyman series is shifted by $\approx-81$ kms$^{-1}$ with respect to that of hydrogen, and becomes visible when the D column density is large enough, albeit this means it is often blended with strong hydrogen lines. The D/H ratio in a metal poor absorption system at $z_{abs}=3.572$ towards QSO PKS1937$-$101 has been re-measured from high resolution, high S/N observations obtained with ESPRESSO\cite{Guarneri2024MNRAS.529..839G} , finding $D/H = 2.6 \pm 0.1 \times 10^{-5}$. This is fully compatible with standard BBN predictions, but in the context of this work is interesting to notice two facts: first, when adding the HIRES spectra to the ESPRESSO and UVES ones, the measurement uncertainty did not improve, indicating that systematic effects associated to differences between various instruments started to dominate. The second fact is that the difference between the ESPRESSO and HIRES spectra, as shown in Fig \ref{fig:deu}, cannot be explained by a constant velocity shift across the plotted wavelength range, but shows how wavelength calibration distortions can vary across a short spectral range. 

A recent attempt to measure carbon isotopic abundance ratio in the IGM was also made more complicated by a cluster of warm pixels falling within the transition of interest. ESPRESSO observations of the QSO J1333$+$1649 taken in 2019, were co-added and a model of two \atom{C}{i} transitions at $z_{abs}=1.776$ was produced to search for the presence of \textsuperscript{13}C in the system, yielding $\cratiomath=34_{-7}^{+10}$. However, the model produced for 2019 observations was found to be incompatible with another set of ESPRESSO observations of the same system taken in 2021, which showed no evidence of $^{13}$C in the system. The right panel of Fig \ref{fig:espresso_warm_pixels} shows the difference between the two epochs coadded spectra.

\subsection{Isotopic ratios in stars}
\begin{figure}
\begin{center}
\begin{tabular}{c}
\includegraphics[height=5.5cm]{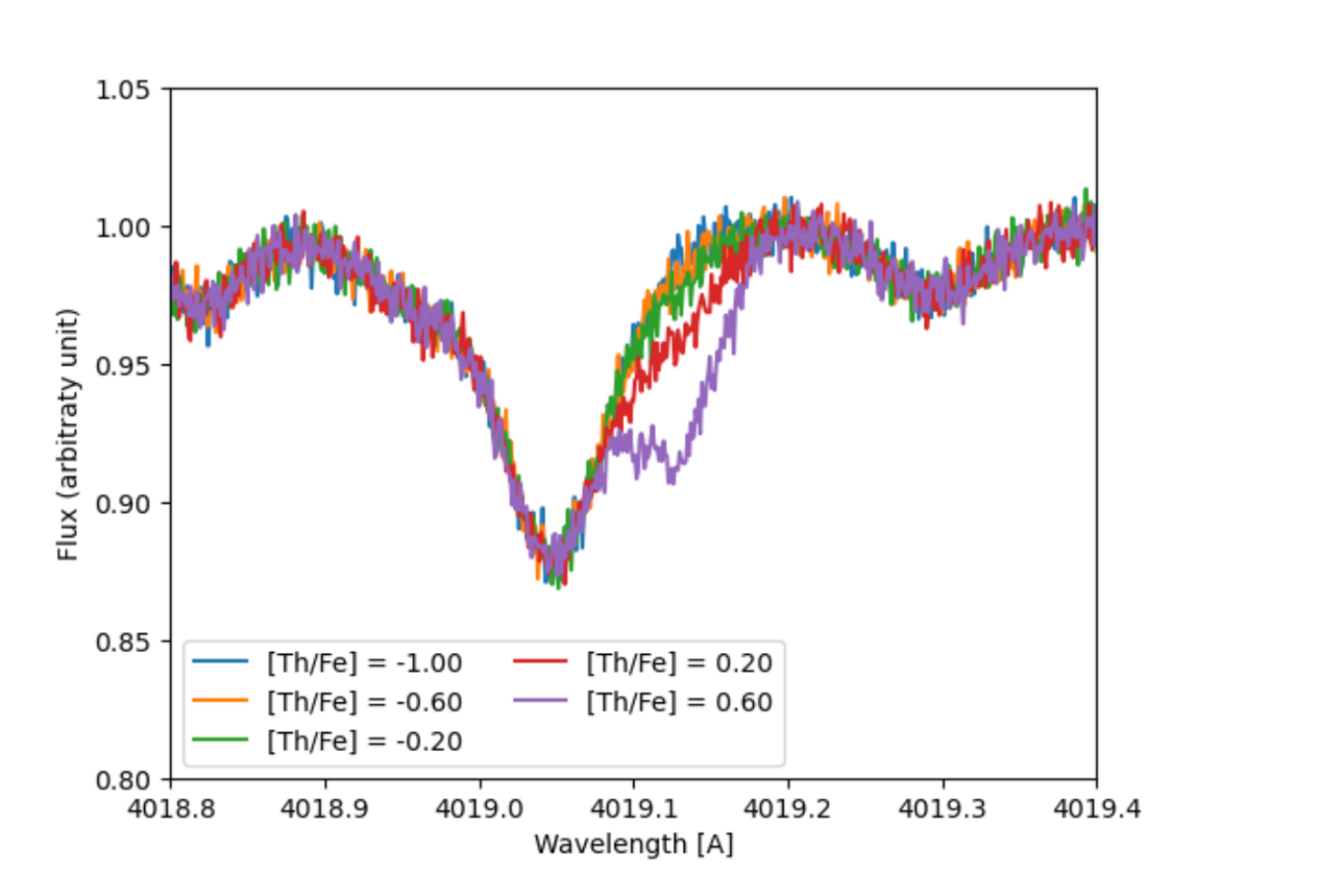}
\end{tabular}
\end{center}
\caption [Synthetic spectrum of a giant star in the Th region]
{ \label{fig:th}
Synthetic spectrum of a giant star in the Th region for several Th abundances. The Th line is tiny and blended with a stronger line. The determination of the Th abundance will require spectral synthesis to reproduce the observed spectrum, and will set stringent requirements on the detector response and on the knowledge of the spectrograph LSF. Taken from Magrini et al.\cite{Magrini2023arXiv231208270M}. } 
\end{figure}

Now that \emph{Gaia} provided good determination of stellar parameters such as luminosity, effective temperature, chemical composition, and surface gravity, the real missing piece of the puzzle is a reliable determination of stellar ages. Cosmochronometry is a very powerful tool for studying stellar evolution, because it is independent of stellar evolution models, and, thanks to its very long decay time, Th is a key element within it. This is why cosmochronometry, and in particular Th/Eu ratios have been chosen as a prominent science case for the proposed new high resolution multi-object facility for the  VLT:  HRMOS\cite{Magrini2023arXiv231208270M} . Fig \ref{fig:th} shows the synthetic Th II 401.9\, nm line for a clump giant, with plotted spectral simulations spanning more than a factor of 10 difference in Th abundance. For the most interesting case of old stars, the Th line is expected to be weak and the surrounding spectral region contains strong V, Cr, Mn, Co, Ni, Ce, and Nd lines. Since the Th line appears in most cases as a blend in the wing of a stronger  absorption line,its abundances can only be retrieved by spectral synthesis of the whole region, reproducing the absorption features by using theoretical (Gaussian or Voigt) profiles, convolved with teh LSF.  Therefore,  the abundance of Th would  be influenced by any effect that causes intrinsically symmetric lines to appear distorted in the spectrum, such as an asymmetric line spread function (LSF), warm pixels, charge transfer effects, and small-scale distortions in the instrument's wavelength calibration (if combining low S/N exposures). Disentangling contributions of Th from the other causes for line asymmetry requires the instrument and detector be to characterised to the same level as is needed for measuring isotopic abundances in the IGM. 

\subsection{Planetary atmospheres}
Certainly, one of the most interesting applications of spectral fidelity to exoplanet studies is the analysis of their atmospheres. The measurement of Oxygen in reflected light of nearby rocky planets has been proposed by coupling extreme adaptive optics to high resolution spectrographs (ANDES \cite{Snellen2015A&A...576A..59S,Marconi2022SPIE12184E..24M} and  METIS\cite{Palle2023arXiv231117075P} at ELT , Ristretto\cite{Lovis2017A&A...599A..16L,Lovis2022SPIE12184E..1QL} at VLT). Despite of a high light rejection factor by the AO systems (the full contrast is expected to be of $\sim$ 10$^{-7}$), the contrast between the exoplanet atmospheric signatures and the stellar spectrum is $<10^{-3}$. The S/N of several tens for the planet's spectrum is expected to be reached after 1 night at the ELT for $\alpha$ Cen or 60 nights on Prox Cen B at the VLT. Should Oxygen be present in the atmospheres of exoplanets, its spectral signature will have to be separated out from those of the host star and the instrument. The latter requires the detector and optical elements contributions to be well understood. Many of the detector requirements overlap with those of the redshift drift experiment (Section \ref{sec:redshift_drift}). 

\section{Reaching Spectral Fidelity}\label{sec:reaching_spectral_fidelity}
In order to reach spectral fidelity, we need three main ingredients: a proper instrument, the ideal calibration system, and an appropriate data reduction. 

\subsection{The Spectrographs}
In the past years, the quest for increasing precision in astronomical spectroscopy has been mostly driven by the development and the requirements of the exoplanet search and the community of astronomers working in this field. The method of exoplanet detection using RV measurements was essentially developed in the 1990s resulting in the seminal paper published in 1995 by Mayor and Queloz\cite{Mayor1995Natur.378..355M}. The basic concept consists in ensuring that the light injection in the spectrograph is independent of the previous light path, and in the adoption of a metrology system that simultaneously and precisely measures RV shifts associated with the spectrograph, effectively removing them. We have now the instruments adequate for reaching unprecedented levels of spectral fidelity, and HARPS can be considered the first of this kind. With the aim to reach 1 ms$^{-1}$ RV precision, HARPS has been developed by improving the intrinsic stability of the spectrograph through high spectral resolution and through an extremely stable fiber illumination. HARPS performance has been improved several times by, e.g., implementing a new guiding system and equipping it with octagonal fibres\cite{LoCurto2015Msngr.162....9L} that have superior scrambling characteristics. The original 1 ms$^{-1}$ RV precision specifications have, thus, been exceeded\cite{LoCurto2017sgvi.confE..23L} . HARPS and its subsequent twins became, through the hundreds of discoveries of Super-Earths and Neptunes over the last 20 years, the new reference for high-precision RV measurements. HARPS has been also the baseline for the development of new generation of spectrographs, such as ESPRESSO aiming at  10 cms$^{-1}$ RV precision, and of a number of other spectrographs worldwide, such as NEID\cite{Lin2022AJ....163..184L} , EXPRES \cite{Blackman2020AJ....159..238B} or KPF \cite{Rubenzahl2023PASP..135l5002R}.

Let's add that the use of atmospheric dispersion correctors, together with octagonal fibres and scrambling are the key components for keeping the spectrograph injection (both in the slit and in the pupil planes) independent of seeing, flux variability, telescope oscillations and position in the sky. While this approach has been shown adequate so far at optical wavelength, it may require improvements in the NIR, because scrambling is less effective, and residual fringing may be present in the science observations (Bouchy et al. 2024, in preparation).

\subsection{The Calibration system}
Metrology, i.e. an accurate and stable calibration system, is the second key element to achieve precise RV measurements and accurate spectroscopy. Wavelength accuracy is obtained by providing two fundamental ingredients: (1) an absolute, broadband and information-rich wavelength source and (2) the ability of the spectrograph and its data reduction procedures to transfer the reference scale (in wavelength or frequency) to a measurement scale (detector pixels) in an accurate manner. Historically this was achieved using Th-Ar hollow-cathode lamps, but these alone are no longer sufficient. A review of the characteristics of the perfect calibrator and the shortcomings of Th-Ar lamps and benefits of astrocombs are described by Murphy et al. \cite{Murphy2007MNRAS.380..839M} . 
Since the current generation of astrocombs are not easy to keep in regular operations and are expensive, some observatories have chosen a mixed solution, making use of Th-Ar lamps in combination with Fabry-Pérot (FP) etalons\cite{Bauer2015A&A...581A.117B} . FP etalons provide evenly-spaced lines, but their wavelengths need to be derived by anchoring to an absolute spectral reference calibrator such as a hollow-cathode lamp. In addition, the separation of the lines has a complex chromatic structure that evolves with time\cite{Terrien2021AJ....161..252T} , and needs to be regularly empirically redetermined. 

\begin{figure}
\begin{center}
\begin{tabular}{c}
\includegraphics[height=7.5cm]{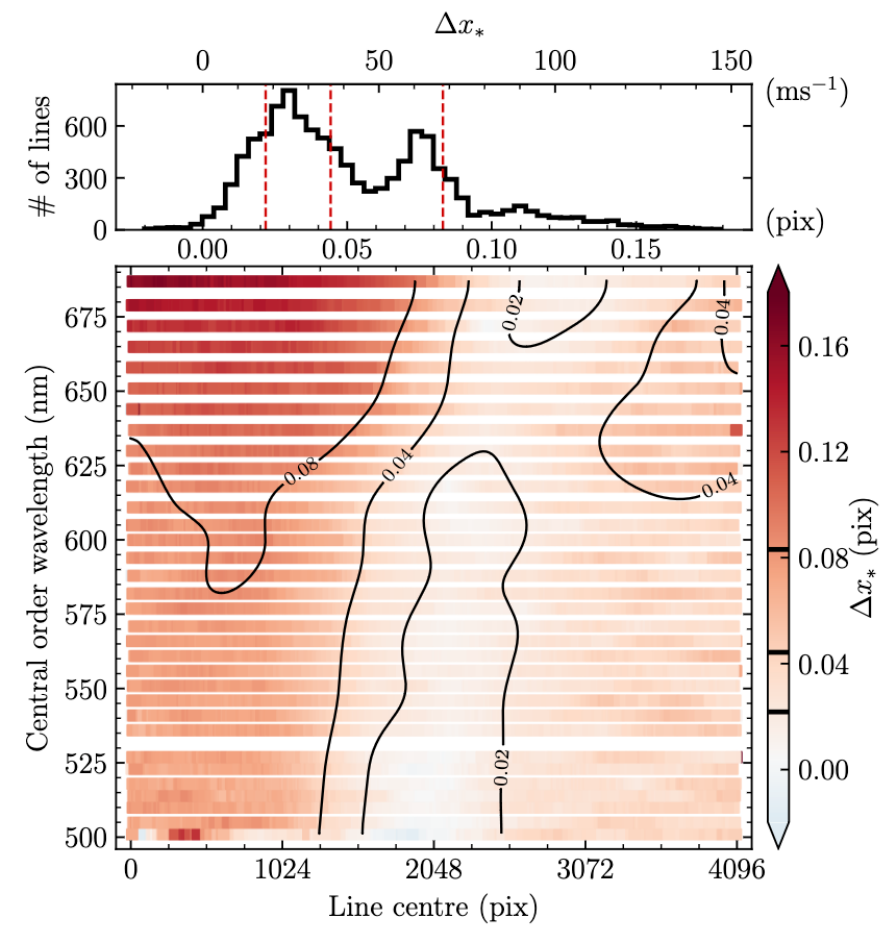}
\end{tabular}
\end{center}
\caption [Impact of using empirical HARPS LSF models on line centroid measurements]
{ \label{fig:psf}
Difference between the center of comb lines as measured with LSF fitting and with Gaussian fitting over a 60\% of the HARPS detector. Taken from Milakovi{\'c} \& Jethwa \cite{Milakovic2024A&A...684A..38M} .} 
\end{figure}

Most importantly, only an astrocomb provides a sufficient number of unblended, infinitely narrow lines to determine the spectrograph's PSF and LSF. PSF modeling has been so far neglected in most scientific analyses, and the calibration lines are fitted with symmetric functions (usually Gaussian), but even in the best spectrographs the PSF is not perfectly symmetric. In addition, the PSF generated by the optical elements becomes more complex because the signal is recorded by a sensor and the PSF is affected by artifacts such as pixelisation, non-linearity, charge transfer inefficiency, brighter-fatter effect, etc. By using astrocomb data, Milakovi{\'c} \& Jethwa \cite{Milakovic2024A&A...684A..38M} measured the HARPS LSF over a large portion of the HARPS detectors. The results are quite interesting: as shown in Fig. \ref{fig:psf}, the centroid of a line fitted using the empirically determined LSF may differ by 100 ms$^{-1}$ with respect to the centroid of the same line determined under the Gaussian LSF assumption. The other interesting finding is that the LSF shape asymmetries are not symmetric with respect to the detector center, but are mostly skewed in the same direction, independently of their location on the CCD. This implies that ignoring LSF variations will induce a systematic shift in the absolute wavelength calibration, in addition to increasing the noise in RV measurements. It also implies that co-adding observations appearing at different areas of the detector (such is the case of transitions with wavelengths in the spectral overlap region of two adjacent echelle orders) should be avoided when aiming for the highest fidelity. 

\subsection{Data Reduction}

The final step to implement spectral fidelity is  an appropriate handling of the data, one that provides an accurate and robust reconstruction of the input astronomical spectrum from the signal recorded by the sensor. One should also eliminate all unwanted spectral signatures, such as those from the instrument, the telescope, and the atmosphere. The decades long experience with exoplanets and the tests with the HARPS astrocomb have shown that the whole chain must be adequately modified to meet the scientific requirements, and this chain includes data reduction software, supported by an appropriate set of calibration frames. Data reduction comprises the software needed to transform the photons detected on pixel at detector coordinates (x,y) into a spectrum calibrated in physical units. It is worth recalling that 1 cms$^{-1}$ corresponds to $\sim$1/80,000\textsuperscript{th} of a pixel on HARPS and $\sim$1/50,000\textsuperscript{th} on ESPRESSO, or 0.2 nm on the detector image plane. 

The standard approach to spectral reduction is based on reverse modeling: given the detector output, the input spectrum is recovered by tracing the signal along the echelle orders and integrating along the transverse direction with an appropriate weighting (“optimal extraction” \cite{Piskunov2002A&A...385.1095P} ). While fast and robust, this method is generally wrong, as it assumes that the instrument PSF is separable along the CCD axes, which is never the case for curved echelle orders. Even improved optimal extraction schemes (e.g.\ Piskunov et al.\cite{Piskunov2021A&A...646A..32P} ) can only mitigate the effects of this issue.
The correct approach (“spectro-perfectionism” or SP \cite{Bolton2010PASP..122..248B,Bolton2012ASPC..461..509B} ), is based on forward modeling, similar to what is commonly applied in other physics experiments:  given the instrument characteristics, the detector output is interpreted as the result of a linear transformation of the input spectrum, which can be recovered by inverting the transformation itself. 
The advantage of SP is indisputable, as it allows a simultaneous modeling of the target signal (including its contaminants, e.g. the sky background) and of the instrumental signature (including a resolution model). Analysis of simulated data by Bolton \& Schlegel \cite{Bolton2010PASP..122..248B} shows that the approach is more than 100 times better than the standard optimal extraction, according to different figures of merit. The cost is an increased technical and computational difficulty in evaluating and inverting huge matrices, a task that was considered intractable at the time  \cite{Bolton2010PASP..122..248B} . Since its original proposal, the convenience and the feasibility of SP has been discussed by several authors (e.g. \cite{Cornachione2019PASP..131l4503C,Guy2023AJ....165..144G} ) and working implementations have been developed for medium-to-high resolution spectrographs\cite{Cornachione2019PASP..131l4503C} . The time is ripe to bring SP to the high fidelity arena. 
The novel approach is feasible in combination with an astrocomb because the latter is a very useful tool to characterize the PSF in each pixel.

\section{Some requirements for spectral fidelity}
Deriving  sensor requirements from astronomical performances is not a trivial task, and this can be done only case by case, deriving properly each subsystem's specifications. They depend on the instrument used and on the science case considered; for instance requirements are different if one wants to use thousands spectral lines of a G-M star for planet search or to analyze a few transitions for the determination of fine structure variability, for measuring the CMB temperature or for measuring isotopic abundances.  We can nevertheless enumerate some of the most general (largely case independent) requirements: 
\begin{enumerate} 

\item Sensors requirements for spectral fidelity are set by the need to obtain precise centroids, manage high contrast and a proper co-adding of (low flux) observations. It is therefore essential to measure and characterize the behaviour of the detector (charge transfer inefficiency, persistency, binary offsets, non-linearity, photo response non-uniformity, low-level warm pixels...) at all relevant flux levels and determine whether they change in time. Measuring intra-pixel sensitivity\cite{Murphy2012MNRAS.422..761M} might be also required. This will require considerable effort, including additional procedures in laboratory (before instrument integration) and a set of appropriate calibrations to be taken at necessary intervals to track any temporal changes in detector properties. Reliably detecting the redshift drift or Oxygen in exoplanet atmospheres will be difficult or even impossible without such advancements.

\item It is imperative to have a  suitable calibration system (e.g., an astrocomb), not only to perform a proper wavelength calibration, but also to be able to recover the PSF and to characterize each pixel used. 
Ideally, the calibration source will have the same input and follow the same optical path of the celestial objects. Preliminary analysis of data collected during an experiment which saw an iodine absorption cell being placed in the ESPRESSO light path, differences of several ms$^{-1}$ between the absolute wavelength calibration scales determined from astrocomb and iodine cell observations were seen. Similar differences are observed between the two image slices of ESPRESSO. Whilst this is work in progress, those results may be associated with light injection and light-path differences, among other effects (T. Schmidt, personal communication). These results further highlight the need to use several independent absolute calibration sources for the most demanding science cases, and that they should be used to cross-check each other as often as necessary. In this context, a proposal for putting an astrocomb in space (as was recently done by F. Pepe, personal communication) is interesting because it would ensure that the calibration light follows the same optical path as science target light. 

\item It is necessary to change how the data are reduced, and move away from spectral extraction to forward modeling using the PSF and other known characteristics of the instrument. This may require, as for many space missions, the full instrument model and the proper detector modeling in the analysis. It is very encouraging to see that software packages to do this are already being developed, such as Pyxel (e.g. PyXel https://esa.gitlab.io/pyxel/, and the contribution by E. George in this volume). For faint object observations, the PSF should be determined from calibrations with similar flux levels. In practice, this will mean taking several exposures at different flux levels, providing information on PSF shape dependency on flux. Once again, astrocombs (or iodine absorption cells, or both) will be required to determine the most appropriate PSF to use for data reduction. Tunable astrocombs \cite{Savchenkov2008PhRvL.101i3902S,Yan2023ApOpt..62.6835Y} could be used to provide additional information on the PSF shape variations on small scales and to measure the intra-pixel response function.

\item All spectral analyses use statistical criteria to fit models and synthetic spectra to the observations, now also using AI. In order to have meaningful results, the flux variance estimates must be correct. There are often hidden hypothesis (e.g.\ that the neighboring pixels are independent from each other) which are not easily tested or sufficiently characterized or reported in the error matrix. Pixel crosstalk, biases, and background noise variations are typical quantities that should be considered. Underestimated flux variances may also be one of the reasons why exceedingly small errors are sometimes associated to spectroscopic results, making them incompatible with other observations, albeit this supposition is not trivial to prove. 

\item Some applications would greatly benefit from zero read-out noise (RON) detectors. One example is exoplanet transit spectroscopy, in which high temporal sampling is required, and zero noise detectors could allow short integrations without penalizing S/N. The other case are observations of  faint sources, that are RON dominated. While, in general, the cases we have described need a final high S/N ratio, we shall consider that such a S/N can be obtained by the combination of many low S/N spectra, so a zero noise detector would translate directly into gain in telescope time.

\item  Since transit spectroscopy requires high observation cadence in a short time, fast read out is another requirement for these observations, as it will translate directly into optimizing telescope time and enhancing the  quality of the in-transit observations. 

\end{enumerate}

\subsection{Shopping list}

We would like to end this contribution emphasizing a couple of sensor requirements that are not related to ``precision astronomy'', but we think are relevant. 

Modern telescopes, either because of their large plate scale, such as the ELTs, or because of their large field of view (e.g. Simonyi Survey Telescope) have huge focal planes. Huge spectroscopic facilities such as  the Widefield Spectroscopic Telescope (WST) \cite{Pasquini2016SPIE.9906E..3CP,Mainieri2024arXiv240305398M} or the MaunaKea Spectroscopic Explorer (MSE) \cite{McConnachie2016arXiv160600043M} , aim to observe many thousands of objects simultaneously,  with hundreds of spectrographs and large sensors.  Pasquini et al.\cite{Pasquini2016SPIE.9906E..3CP} show that {\it curved detectors} can provide significant simplification to the spectrographs' design and optics, decreasing their costs and enhancing their performances. They seem a must for this new generation of telescopes, and several developments are ongoing with encouraging results\cite{Hugot2019SPIE11180E..2YH,OMasta2022SPIE12107E..1SO} . 

The need for many detectors also implies that costs of new ground based facilities become prohibitive, and already now  the high cost of sensors limit the development of new instruments  that are simply not affordable. 

As a result of needing extremely precise knowledge of our sensors at each pixel -- and potentially even at sub-pixel scales -- a significant number of detailed and time-intensive calibrations will be required. It seems therefore essential that the detector system characteristics are stable in time, because it might be prohibitive to repeat often huge calibration sets. On the other hand, new spectrographs should consider to have detector calibration systems in situ, to be repeated in standard operations. 

\subsection*{Disclosures}
Authors declare no financial interests in the manuscript and no other potential conflicts of interest.

\subsection*{Code, Data, and Materials Availability}
All figures show publicly available datasets or have been reproduced from other works. Raw data used in this work can be accessed from ESO and Keck archives. Processed data will be made available upon reasonable request to the authors.

%%%%% References %%%%%

\bibliography{export-bibtex}   % bibliography data in report.bib

\begin{thebibliography}{10}

\bibitem{Euclid2020A&A...635A.139E}
{Euclid Collaboration}, P.~{Paykari}, T.~{Kitching}, {\em et~al.}, ``{Euclid
  preparation. VI. Verifying the performance of cosmic shear experiments},''
  {\em \aap} {\bf 635}, A139  (2020).

\bibitem{Prusti2016A&A...595A...1G}
{Gaia Collaboration}, T.~{Prusti}, J.~H.~J. {de Bruijne}, {\em et~al.}, ``{The
  Gaia mission},'' {\em \aap} {\bf 595}, A1  (2016).

\bibitem{Davies2021Msngr.182...17D}
R.~{Davies}, V.~{H{\"o}rmann}, S.~{Rabien}, {\em et~al.}, ``{MICADO: The
  Multi-Adaptive Optics Camera for Deep Observations},'' {\em The Messenger}
  {\bf 182}, 17--21  (2021).

\bibitem{Bergemann2021MNRAS.508.2236B}
M.~{Bergemann}, R.~{Hoppe}, E.~{Semenova}, {\em et~al.}, ``{Solar oxygen
  abundance},'' {\em \mnras} {\bf 508}, 2236--2253  (2021).

\bibitem{Webb2024MNRAS.528.6550W}
J.~K. {Webb} and C.-C. {Lee}, ``{Convergence properties of fine structure
  constant measurements using quasar absorption systems},'' {\em \mnras} {\bf
  528}, 6550--6558  (2024).

\bibitem{Lee2021MNRAS.507...27L_nonunique}
C.-C. {Lee}, J.~K. {Webb}, D.~{Milakovi{\'c}}, {\em et~al.}, ``{Non-uniqueness
  in quasar absorption models and implications for measurements of the fine
  structure constant},'' {\em \mnras} {\bf 507}, 27--42  (2021).

\bibitem{Pepe2021A&A...645A..96P}
F.~{Pepe}, S.~{Cristiani}, R.~{Rebolo}, {\em et~al.}, ``{ESPRESSO at VLT.
  On-sky performance and first results},'' {\em \aap} {\bf 645}, A96  (2021).

\bibitem{Schindler2021ApJ...906...12S}
J.-T. {Schindler}, X.~{Fan}, M.~{Novak}, {\em et~al.}, ``{A Closer Look at Two
  of the Most Luminous Quasars in the Universe},'' {\em \apj} {\bf 906}, 12
  (2021).

\bibitem{Cristiani2023MNRAS.522.2019C}
S.~{Cristiani}, M.~{Porru}, F.~{Guarneri}, {\em et~al.}, ``{Spectroscopy of
  QUBRICS quasar candidates: 1672 new redshifts and a golden sample for the
  Sandage test of the redshift drift},'' {\em \mnras} {\bf 522}, 2019--2028
  (2023).

\bibitem{Sandage1962ApJ...136..319S}
A.~{Sandage}, ``{The Change of Redshift and Apparent Luminosity of Galaxies due
  to the Deceleration of Selected Expanding Universes.},'' {\em \apj} {\bf
  136}, 319  (1962).

\bibitem{McVittie1962ApJ...136..334M}
G.~C. {McVittie}, ``{Appendix to The Change of Redshift and Apparent Luminosity
  of Galaxies due to the Deceleration of Selected Expanding Universes.},'' {\em
  \apj} {\bf 136}, 334  (1962).

\bibitem{Loeb1998ApJ...499L.111L}
A.~{Loeb}, ``{Direct Measurement of Cosmological Parameters from the Cosmic
  Deceleration of Extragalactic Objects},'' {\em \apjl} {\bf 499}, L111--L114
  (1998).

\bibitem{Liske2008MNRAS.386.1192L}
J.~{Liske}, A.~{Grazian}, E.~{Vanzella}, {\em et~al.}, ``{Cosmic dynamics in
  the era of Extremely Large Telescopes},'' {\em \mnras} {\bf 386}, 1192--1218
  (2008).

\bibitem{Mayor2003Msngr.114...20M}
M.~{Mayor}, F.~{Pepe}, D.~{Queloz}, {\em et~al.}, ``{Setting New Standards with
  HARPS},'' {\em The Messenger} {\bf 114}, 20--24  (2003).

\bibitem{Milakovic2020MNRAS.493.3997M}
D.~{Milakovi{\'c}}, L.~{Pasquini}, J.~K. {Webb}, {\em et~al.}, ``{Precision and
  consistency of astrocombs},'' {\em \mnras} {\bf 493}, 3997--4011  (2020).

\bibitem{Uzan2011LRR....14....2U}
J.-P. {Uzan}, ``{Varying Constants, Gravitation and Cosmology},'' {\em Living
  Reviews in Relativity} {\bf 14}, 2  (2011).

\bibitem{Uzan2024arXiv241007281U}
J.-P. {Uzan}, ``{Fundamental constants: from measurement to the universe, a
  window on gravitation and cosmology},'' {\em arXiv e-prints} ,
  arXiv:2410.07281  (2024).

\bibitem{Bekenstein1982PhRvD..25.1527B}
J.~D. {Bekenstein}, ``{Fine-structure constant: Is it really a constant?},''
  {\em \prd} {\bf 25}, 1527--1539  (1982).

\bibitem{Barrow2012PhRvD..85b3514B}
J.~D. {Barrow} and S.~Z.~W. {Lip}, ``{Generalized theory of varying alpha},''
  {\em \prd} {\bf 85}, 023514  (2012).

\bibitem{Shaw2005PhRvD..71f3525S}
D.~J. {Shaw} and J.~D. {Barrow}, ``{Varying couplings in electroweak theory},''
  {\em \prd} {\bf 71}, 063525  (2005).

\bibitem{Sandvik2002PhRvL..88c1302B}
H.~B. {Sandvik}, J.~D. {Barrow}, and J.~{Magueijo}, ``{A Simple Cosmology with
  a Varying Fine Structure Constant},'' {\em \prl} {\bf 88}, 031302  (2002).

\bibitem{Mota2004PhLB..581..141M}
D.~F. {Mota} and J.~D. {Barrow}, ``{Varying alpha in a more realistic
  universe},'' {\em Physics Letters B} {\bf 581}, 141--146  (2004).

\bibitem{Rosenband2008Sci...319.1808R}
T.~{Rosenband}, D.~B. {Hume}, P.~O. {Schmidt}, {\em et~al.}, ``{Frequency Ratio
  of Al$^{+}$ and Hg$^{+}$ Single-Ion Optical Clocks; Metrology at the 17th
  Decimal Place},'' {\em Science} {\bf 319}, 1808  (2008).

\bibitem{Dzuba1999PhRvA..59..230D}
V.~A. {Dzuba}, V.~V. {Flambaum}, and J.~K. {Webb}, ``{Calculations of the
  relativistic effects in many-electron atoms and space-time variation of
  fundamental constants},'' {\em \pra} {\bf 59}, 230--237  (1999).

\bibitem{Dzuba1999PhRvL..82..888D}
V.~A. {Dzuba}, V.~V. {Flambaum}, and J.~K. {Webb}, ``{Space-Time Variation of
  Physical Constants and Relativistic Corrections in Atoms},'' {\em \prl} {\bf
  82}, 888--891  (1999).

\bibitem{Webb1999PhRvL..82..884W}
J.~K. {Webb}, V.~V. {Flambaum}, C.~W. {Churchill}, {\em et~al.}, ``{Search for
  Time Variation of the Fine Structure Constant},'' {\em \prl} {\bf 82},
  884--887  (1999).

\bibitem{Dekker2000SPIE.4008..534D}
H.~{Dekker}, S.~{D'Odorico}, A.~{Kaufer}, {\em et~al.}, ``{Design,
  construction, and performance of UVES, the echelle spectrograph for the UT2
  Kueyen Telescope at the ESO Paranal Observatory},'' in {\em Optical and IR
  Telescope Instrumentation and Detectors},  M.~{Iye} and A.~F. {Moorwood},
  Eds., {\em Society of Photo-Optical Instrumentation Engineers (SPIE)
  Conference Series} {\bf 4008}, 534--545  (2000).

\bibitem{Vogt1994SPIE.2198..362V}
S.~S. {Vogt}, S.~L. {Allen}, B.~C. {Bigelow}, {\em et~al.}, ``{HIRES: the
  high-resolution echelle spectrometer on the Keck 10-m Telescope},'' in {\em
  Instrumentation in Astronomy VIII},  D.~L. {Crawford} and E.~R. {Craine},
  Eds., {\em Society of Photo-Optical Instrumentation Engineers (SPIE)
  Conference Series} {\bf 2198}, 362  (1994).

\bibitem{Webb2001PhRvL..87i1301W}
J.~K. {Webb}, M.~T. {Murphy}, V.~V. {Flambaum}, {\em et~al.}, ``{Further
  Evidence for Cosmological Evolution of the Fine Structure Constant},'' {\em
  \prl} {\bf 87}, 091301  (2001).

\bibitem{Webb2003Ap&SS.283..565W}
J.~K. {Webb}, M.~T. {Murphy}, V.~V. {Flambaum}, {\em et~al.}, ``{Does the fine
  structure constant vary? A third quasar absorption sample consistent with
  varying {\ensuremath{\alpha}}},'' {\em \apss} {\bf 283}, 565--575  (2003).

\bibitem{Murphy2003MNRAS.345..609M_measurements}
M.~T. {Murphy}, J.~K. {Webb}, and V.~V. {Flambaum}, ``{Further evidence for a
  variable fine-structure constant from Keck/HIRES QSO absorption spectra},''
  {\em \mnras} {\bf 345}, 609--638  (2003).

\bibitem{Webb2011PhRvL.107s1101W}
J.~K. {Webb}, J.~A. {King}, M.~T. {Murphy}, {\em et~al.}, ``{Indications of a
  Spatial Variation of the Fine Structure Constant},'' {\em \prl} {\bf 107},
  191101  (2011).

\bibitem{King2012MNRAS.422.3370K}
J.~A. {King}, J.~K. {Webb}, M.~T. {Murphy}, {\em et~al.}, ``{Spatial variation
  in the fine-structure constant - new results from VLT/UVES},'' {\em \mnras}
  {\bf 422}, 3370--3414  (2012).

\bibitem{Griest2010ApJ...708..158G}
K.~{Griest}, J.~B. {Whitmore}, A.~M. {Wolfe}, {\em et~al.}, ``{Wavelength
  Accuracy of the Keck HIRES Spectrograph and Measuring Changes in the Fine
  Structure Constant},'' {\em \apj} {\bf 708}, 158--170  (2010).

\bibitem{Whitmore2010ApJ...723...89W}
J.~B. {Whitmore}, M.~T. {Murphy}, and K.~{Griest}, ``{Wavelength Calibration of
  the VLT-UVES Spectrograph},'' {\em \apj} {\bf 723}, 89--99  (2010).

\bibitem{Whitmore2015MNRAS.447..446W}
J.~B. {Whitmore} and M.~T. {Murphy}, ``{Impact of instrumental systematic
  errors on fine-structure constant measurements with quasar spectra},'' {\em
  \mnras} {\bf 447}, 446--462  (2015).

\bibitem{Rahmani2013MNRAS.435..861R}
H.~{Rahmani}, M.~{Wendt}, R.~{Srianand}, {\em et~al.}, ``{The UVES large
  program for testing fundamental physics - II. Constraints on a change in
  {\ensuremath{\mu}} towards quasar HE 0027-1836},'' {\em \mnras} {\bf 435},
  861--878  (2013).

\bibitem{Molaro2008A&A...481..559M}
P.~{Molaro}, S.~A. {Levshakov}, S.~{Monai}, {\em et~al.}, ``{UVES radial
  velocity accuracy from asteroid observations. I. Implications for fine
  structure constant variability},'' {\em \aap} {\bf 481}, 559--569  (2008).

\bibitem{Dumont2017MNRAS.468.1568D}
V.~{Dumont} and J.~K. {Webb}, ``{Modelling long-range wavelength distortions in
  quasar absorption echelle spectra},'' {\em \mnras} {\bf 468}, 1568--1574
  (2017).

\bibitem{Milakovic2021MNRAS.500....1M}
D.~{Milakovi{\'c}}, C.-C. {Lee}, R.~F. {Carswell}, {\em et~al.}, ``{A new era
  of fine structure constant measurements at high redshift},'' {\em \mnras}
  {\bf 500}, 1--21  (2021).

\bibitem{Milakovic2024A&A...684A..38M}
D.~{Milakovi{\'c}} and P.~{Jethwa}, ``{A new method for instrumental profile
  reconstruction of high-resolution spectrographs},'' {\em \aap} {\bf 684}, A38
   (2024).

\bibitem{Schmidt2021A&A...646A.144S}
T.~M. {Schmidt}, P.~{Molaro}, M.~T. {Murphy}, {\em et~al.}, ``{Fundamental
  physics with ESPRESSO: Towards an accurate wavelength calibration for a
  precision test of the fine-structure constant},'' {\em \aap} {\bf 646}, A144
  (2021).

\bibitem{Milakovic2023MmSAI..94b.270M}
D.~{Milakovi{\'c}}, C.~C. {Lee}, P.~{Molaro}, {\em et~al.}, ``{Methods for
  quasar absorption system measurements of the fine structure constant in the
  2020s and beyond},'' in {\em Memorie della Societa Astronomica Italiana},
  {\bf 94}, 270  (2023).

\bibitem{Guarneri2024MNRAS.529..839G}
F.~{Guarneri}, L.~{Pasquini}, V.~{D'Odorico}, {\em et~al.}, ``{Fundamental
  physics with ESPRESSO: a new determination of the D/H ratio towards
  PKS1937-101},'' {\em \mnras} {\bf 529}, 839--854  (2024).

\bibitem{Magrini2023arXiv231208270M}
L.~{Magrini}, T.~{Bensby}, A.~{Brucalassi}, {\em et~al.}, ``{HRMOS White Paper:
  Science Motivation},'' {\em arXiv e-prints} , arXiv:2312.08270  (2023).

\bibitem{Snellen2015A&A...576A..59S}
I.~{Snellen}, R.~{de Kok}, J.~L. {Birkby}, {\em et~al.}, ``{Combining
  high-dispersion spectroscopy with high contrast imaging: Probing rocky
  planets around our nearest neighbors},'' {\em \aap} {\bf 576}, A59  (2015).

\bibitem{Marconi2022SPIE12184E..24M}
A.~{Marconi}, M.~{Abreu}, V.~{Adibekyan}, {\em et~al.}, ``{ANDES, the high
  resolution spectrograph for the ELT: science case, baseline design and path
  to construction},'' in {\em Ground-based and Airborne Instrumentation for
  Astronomy IX},  C.~J. {Evans}, J.~J. {Bryant}, and K.~{Motohara}, Eds., {\em
  Society of Photo-Optical Instrumentation Engineers (SPIE) Conference Series}
  {\bf 12184}, 1218424  (2022).

\bibitem{Palle2023arXiv231117075P}
E.~{Palle}, K.~{Biazzo}, E.~{Bolmont}, {\em et~al.}, ``{Ground-breaking
  Exoplanet Science with the ANDES spectrograph at the ELT},'' {\em arXiv
  e-prints} , arXiv:2311.17075  (2023).

\bibitem{Lovis2017A&A...599A..16L}
C.~{Lovis}, I.~{Snellen}, D.~{Mouillet}, {\em et~al.}, ``{Atmospheric
  characterization of Proxima b by coupling the SPHERE high-contrast imager to
  the ESPRESSO spectrograph},'' {\em \aap} {\bf 599}, A16  (2017).

\bibitem{Lovis2022SPIE12184E..1QL}
C.~{Lovis}, N.~{Blind}, B.~{Chazelas}, {\em et~al.}, ``{RISTRETTO:
  high-resolution spectroscopy at the diffraction limit of the VLT},'' in {\em
  Ground-based and Airborne Instrumentation for Astronomy IX},  C.~J. {Evans},
  J.~J. {Bryant}, and K.~{Motohara}, Eds., {\em Society of Photo-Optical
  Instrumentation Engineers (SPIE) Conference Series} {\bf 12184}, 121841Q
  (2022).

\bibitem{Mayor1995Natur.378..355M}
M.~{Mayor} and D.~{Queloz}, ``{A Jupiter-mass companion to a solar-type
  star},'' {\em \nat} {\bf 378}, 355--359  (1995).

\bibitem{LoCurto2015Msngr.162....9L}
G.~{Lo Curto}, F.~{Pepe}, G.~{Avila}, {\em et~al.}, ``{HARPS Gets New Fibres
  After 12 Years of Operations},'' {\em The Messenger} {\bf 162}, 9--15
  (2015).

\bibitem{LoCurto2017sgvi.confE..23L}
G.~{Lo Curto}, ``{Instrument Talk : HARPS},'' in {\em ESO Calibration Workshop:
  The Second Generation VLT Instruments and Friends},  23  (2017).

\bibitem{Lin2022AJ....163..184L}
A.~S.~J. {Lin}, A.~{Monson}, S.~{Mahadevan}, {\em et~al.}, ``{Observing the Sun
  as a Star: Design and Early Results from the NEID Solar Feed},'' {\em \aj}
  {\bf 163}, 184  (2022).

\bibitem{Blackman2020AJ....159..238B}
R.~T. {Blackman}, D.~A. {Fischer}, C.~A. {Jurgenson}, {\em et~al.},
  ``{Performance Verification of the EXtreme PREcision Spectrograph},'' {\em
  \aj} {\bf 159}, 238  (2020).

\bibitem{Rubenzahl2023PASP..135l5002R}
R.~A. {Rubenzahl}, S.~{Halverson}, J.~{Walawender}, {\em et~al.}, ``{Staring at
  the Sun with the Keck Planet Finder: An Autonomous Solar Calibrator for High
  Signal-to-noise Sun-as-a-star Spectra},'' {\em \pasp} {\bf 135}, 125002
  (2023).

\bibitem{Murphy2007MNRAS.380..839M}
M.~T. {Murphy}, T.~{Udem}, R.~{Holzwarth}, {\em et~al.}, ``{High-precision
  wavelength calibration of astronomical spectrographs with laser frequency
  combs},'' {\em \mnras} {\bf 380}, 839--847  (2007).

\bibitem{Bauer2015A&A...581A.117B}
F.~F. {Bauer}, M.~{Zechmeister}, and A.~{Reiners}, ``{Calibrating echelle
  spectrographs with Fabry-P{\'e}rot etalons},'' {\em \aap} {\bf 581}, A117
  (2015).

\bibitem{Terrien2021AJ....161..252T}
R.~C. {Terrien}, J.~P. {Ninan}, S.~A. {Diddams}, {\em et~al.}, ``{Broadband
  Stability of the Habitable Zone Planet Finder Fabry-P{\'e}rot Etalon
  Calibration System: Evidence for Chromatic Variation},'' {\em \aj} {\bf 161},
  252  (2021).

\bibitem{Piskunov2002A&A...385.1095P}
N.~E. {Piskunov} and J.~A. {Valenti}, ``{New algorithms for reducing
  cross-dispersed echelle spectra},'' {\em \aap} {\bf 385}, 1095--1106  (2002).

\bibitem{Piskunov2021A&A...646A..32P}
N.~{Piskunov}, A.~{Wehrhahn}, and T.~{Marquart}, ``{Optimal extraction of
  echelle spectra: Getting the most out of observations},'' {\em \aap} {\bf
  646}, A32  (2021).

\bibitem{Bolton2010PASP..122..248B}
A.~S. {Bolton} and D.~J. {Schlegel}, ``{Spectro-Perfectionism: An Algorithmic
  Framework for Photon Noise-Limited Extraction of Optical Fiber
  Spectroscopy},'' {\em \pasp} {\bf 122}, 248  (2010).

\bibitem{Bolton2012ASPC..461..509B}
A.~S. {Bolton}, S.~{Bailey}, J.~{Brownstein}, {\em et~al.}, ``{What is a
  Spectrum?},'' in {\em Astronomical Data Analysis Software and Systems XXI},
  P.~{Ballester}, D.~{Egret}, and N.~P.~F. {Lorente}, Eds., {\em Astronomical
  Society of the Pacific Conference Series} {\bf 461}, 509  (2012).

\bibitem{Cornachione2019PASP..131l4503C}
M.~A. {Cornachione}, A.~S. {Bolton}, J.~D. {Eastman}, {\em et~al.}, ``{A Full
  Implementation of Spectro-perfectionism for Precise Radial Velocity Exoplanet
  Detection: A Test Case With the MINERVA Reduction Pipeline},'' {\em \pasp}
  {\bf 131}, 124503  (2019).

\bibitem{Guy2023AJ....165..144G}
J.~{Guy}, S.~{Bailey}, A.~{Kremin}, {\em et~al.}, ``{The Spectroscopic Data
  Processing Pipeline for the Dark Energy Spectroscopic Instrument},'' {\em
  \aj} {\bf 165}, 144  (2023).

\bibitem{Murphy2012MNRAS.422..761M}
M.~T. {Murphy}, C.~R. {Locke}, P.~S. {Light}, {\em et~al.}, ``{Laser frequency
  comb techniques for precise astronomical spectroscopy},'' {\em \mnras} {\bf
  422}, 761--771  (2012).

\bibitem{Savchenkov2008PhRvL.101i3902S}
A.~A. {Savchenkov}, A.~B. {Matsko}, V.~S. {Ilchenko}, {\em et~al.}, ``{Tunable
  Optical Frequency Comb with a Crystalline Whispering Gallery Mode
  Resonator},'' {\em \prl} {\bf 101}, 093902  (2008).

\bibitem{Yan2023ApOpt..62.6835Y}
J.~{Yan}, Y.~{Wang}, and H.~{Zeng}, ``{Generation of GHz line-spacing tunable
  optical frequency combs using Talbot effects},'' {\em \ao} {\bf 62}, 6835
  (2023).

\bibitem{Pasquini2016SPIE.9906E..3CP}
L.~{Pasquini}, B.~{Delabre}, R.~{Ellis}, {\em et~al.}, ``{New telescope designs
  suitable for massively multiplexed spectroscopy},'' in {\em Ground-based and
  Airborne Telescopes VI},  H.~J. {Hall}, R.~{Gilmozzi}, and H.~K. {Marshall},
  Eds., {\em Society of Photo-Optical Instrumentation Engineers (SPIE)
  Conference Series} {\bf 9906}, 99063C  (2016).

\bibitem{Mainieri2024arXiv240305398M}
V.~{Mainieri}, R.~I. {Anderson}, J.~{Brinchmann}, {\em et~al.}, ``{The
  Wide-field Spectroscopic Telescope (WST) Science White Paper},'' {\em arXiv
  e-prints} , arXiv:2403.05398  (2024).

\bibitem{McConnachie2016arXiv160600043M}
A.~{McConnachie}, C.~{Babusiaux}, M.~{Balogh}, {\em et~al.}, ``{The Detailed
  Science Case for the Maunakea Spectroscopic Explorer: the Composition and
  Dynamics of the Faint Universe},'' {\em arXiv e-prints} , arXiv:1606.00043
  (2016).

\bibitem{Hugot2019SPIE11180E..2YH}
E.~{Hugot}, S.~{Lombardo}, T.~{Behaghel}, {\em et~al.}, ``{Curved sensors:
  experimental performance of CMOS prototypes and wide field related
  imagers},'' in {\em International Conference on Space Optics \&mdash; ICSO
  2018},  Z.~{Sodnik}, N.~{Karafolas}, and B.~{Cugny}, Eds., {\em Society of
  Photo-Optical Instrumentation Engineers (SPIE) Conference Series} {\bf
  11180}, 111802Y  (2019).

\bibitem{OMasta2022SPIE12107E..1SO}
M.~R. {O'Masta}, B.-M. {Nguyen}, A.~{Gurga}, {\em et~al.}, ``{Curving of
  large-format infrared sensors},'' in {\em Infrared Technology and
  Applications XLVIII},  B.~F. {Andresen}, G.~F. {Fulop}, L.~{Zheng}, {\em
  et~al.}, Eds., {\em Society of Photo-Optical Instrumentation Engineers (SPIE)
  Conference Series} {\bf 12107}, 121071S  (2022).

\end{thebibliography}
\bibliographystyle{spiejour}   % makes bibtex use spiejour.bst

\end{spacing}
\end{document}